\documentclass{ar2e}
\usepackage{psfrag}
\usepackage[sort]{natbib}
\bibpunct{(}{)}{,}{a}{}{,}
\newcommand{\vomega}{\mbox{\boldmath $\omega$}}

\begin{document}

\input epsf.tex    

\input psfig.sty

\jname{Annual Review of Fluid Mechanics}
\jyear{2011}
\jvol{}
\ARinfo{1056-8700/97/0610-00}

\title{Scale Interactions in Magnetohydrodynamic Turbulence}

\markboth{Mininni}{Scale Interactions in MHD Turbulence}

\author{Pablo D. Mininni
\affiliation{Departamento de F\'{\i}sica, Facultad de Ciencias Exactas 
    y Naturales, Universidad de Buenos Aires and CONICET, Ciudad 
    Universitaria, 1428 Buenos Aires, Argentina, \\
    and \\
    NCAR, P.O.~Box 3000, Boulder, Colorado 80307-3000, USA.}}


\begin{abstract}
This article reviews recent studies of scale interactions in 
magnetohydrodynamic turbulence. The present day increase of computing power, 
which allows for the exploration of different configurations of turbulence 
in conducting flows, and the development of shell-to-shell transfer 
functions, has led to detailed studies of interactions between the velocity 
and the magnetic field and between scales. In particular, 
processes such as induction and dynamo action, the damping of velocity 
fluctuations by the Lorentz force, or the development of anisotropies, 
can be characterized at different scales. In this context we consider 
three different configurations often studied in the literature: 
mechanically forced turbulence, freely decaying turbulence, and 
turbulence in the presence of a uniform magnetic field. Each 
configuration is of interest for different geophysical and 
astrophysical applications. Local and non-local transfers are 
discussed for each case. While the transfer between scales 
of solely kinetic or solely magnetic energy is local, transfers 
between kinetic and magnetic fields are observed to be local or 
non-local depending on the configuration. Scale interactions in the 
cascade of magnetic helicity are also reviewed. Based on the 
results, the validity of several usual assumptions in hydrodynamic 
turbulence, such as isotropy of the small scales or universality, is 
discussed.
\end{abstract}

\maketitle

\section{Introduction}

Turbulence is a multiscale phenomenon ubiquitous in geophysical and 
astrophysical flows. In many of these flows, the coupling of a conducting 
fluid with electromagnetic fields requires consideration of the 
magnetohydrodynamic (MHD) equations (see, e.g., \citealt{Moffatt}). 
The equations describe the dynamics of non-relativistic 
conducting fluids as, e.g., in the Earth's core or in industrial 
applications, and under some approximations they can also describe the 
large-scale behavior of magnetospheric, space, and astrophysical plasmas. 
In these latter cases, care must be taken to consider only the scales where 
a one-fluid approximation holds, as scales small enough may require 
consideration of kinetic plasma effects such as ambipolar diffusion in 
weakly ionized plasmas as the interstellar medium, or the Hall current for 
highly ionized media such as small scales in the solar wind. However, 
in those cases the MHD equations still give a good description of large 
scales, and the approximation gives a useful approach to get 
lowest-order physical insight into the fate of the flows.

In the simplest case, that of an incompressible flow with constant mass 
density, the equations give the evolution of the bulk fluid velocity 
${\bf u}$ and of the magnetic field ${\bf b}$:
\begin{equation}
\partial_t {\bf u} + {\bf u}\cdot \nabla {\bf u} = - \nabla p +
{\bf b}\cdot \nabla {\bf b} + \nu \nabla^2 {\bf u} ,
\label{eq:momentum}
\end{equation}
\begin{equation}
\partial_t {\bf b} + {\bf u}\cdot \nabla {\bf b} = {\bf b}\cdot \nabla {\bf u}
+ \eta \nabla^2 {\bf b} ,
\label{eq:Induction}
\end{equation}
where the magnetic field is written in Alfvenic units, the density is 
set to unity, and $p$ is the (fluid plus magnetic) pressure. The 
kinematic viscosity $\nu$ and magnetic diffusivity $\eta$ 
control respectively the viscous and 
Ohmic dissipation. These equations are constrained by the 
incompressibility condition and by the solenoidal character of the 
magnetic field,
\begin{equation}
\nabla \cdot {\bf u} =0, \,\,\, \nabla \cdot {\bf b} =0 .
\label{eq:Incompresible}
\end{equation}

Two different Reynolds numbers can be defined in MHD flows. The mechanical 
Reynolds number
\begin{equation}
Re = \frac{UL}{\nu} ,
\end{equation}
which is the ratio of convective to viscous forces (where $U$ is the rms 
velocity and $L$ a characteristic lengthscale of the flow), and the 
magnetic Reynolds number
\begin{equation}
Rm = \frac{UL}{\eta} ,
\end{equation}
that can be interpreted as the ratio of induction to Ohmic dissipation. In 
many flows these Reynolds numbers are very large, and the flows are in a 
turbulent regime.

While in the hydrodynamic case the phenomenological theory of Kolmogorov 
(K41) gives to a good approximation (albeit intermittency corrections) the 
power law of the energy spectrum, no clearly established phenomenological 
counterpart exists in MHD. This has many 
implications as the energy dissipation rate (required to predict, e.g., 
heating rates in solar and space physics) depends on the slope of the 
energy spectrum. Also, subgrid models, required to do numerical modeling 
in astrophysics and geophysics given the large scale separation 
involved in such flows, are less developed in MHD as a result 
of the lack of detailed knowledge of its energy spectrum.

In the Kolmogorov description of hydrodynamic turbulence, the interactions 
of similar size eddies play the basic role of cascading the energy to 
smaller scales on a scale-dependent time scale $\tau_\ell \sim \ell/u_\ell$, 
where $\ell$ is the examined length scale and $u_\ell$ the characteristic 
velocity at that scale. This time scale, which is proportional to the eddy 
turnover time at the scale $\ell$, is the only time scale available on 
dimensional grounds in the inertial range, provided enough scale separation 
exists between forcing and dissipation. In this context, interactions 
between scales are local (in spectral space) as dominant interactions are 
between eddies of similar sizes. One then expects the statistical properties 
of sufficiently small scales to be independent of the way turbulence is 
generated, and to have therefore universal character. Recent experiments 
showed deviations from this behavior even for simple hydrodynamic flows 
(e.g., slower than expected recovery of isotropy, or presence of long-time 
correlations in the small scales; see 
\citealt{Wiltse93,Wiltse98,Shen00,Carlier01,Poulain06}). Numerical 
simulations also gave evidence of the presence of non-local interactions 
with the large scale flow playing a role in the cascade of energy 
\citep{Domaradzki88,Domaradzki90,Zhou93,Alexakis05b}. In numerical 
simulations with Reynolds numbers as high as $R_\lambda \approx 800$, it was 
observed that 20\% of the energy flux in the small scales was due to 
interactions with the large scale flow \citep{Mininni06}. However, more 
recent simulations with Reynolds numbers up to $R_\lambda \approx 1300$ 
using spatial resolutions of $2048^3$ grid points showed that as the 
Reynolds number is increased, the percentage of the non-local flux 
decreases as a power law of the Reynolds number, suggesting that the flux 
in hydrodynamic turbulence may be predominantly local for very large 
Reynolds number \citep{Mininni08}. Recent theoretical results put this in 
firmer grounds \citep{Eyink09,Aluie09}, showing that the energy flux in 
hydrodynamic turbulence is local in the limit of infinite Reynolds 
number and obtaining bounds on the scaling of the non-local 
contribution to the flux with Reynolds number which are in agreement 
with the numerical results.

The case for MHD turbulence is less clear and has given rise to more 
controversies. Several attempts have been done to extend the 
phenomenological arguments of Kolmogorov to conducting flows (see, e.g., 
\citealt{Iroshnikov63,Kraichnan65,Matthaeus89,Goldreich95,Boldyrev06}). 
However, the MHD equivalent of the 4/5 law in hydrodynamic turbulence 
(the Politano-Pouquet relations, see \citealt{Politano98a,Politano98b}) 
couple the velocity and the magnetic field in a way that can be 
compatible with several power law behaviors; in three dimensions they 
read
\begin{equation}
\left< \delta z_\parallel^\mp ({\bf l}) \left|\delta {\bf z}^\pm ({\bf l})
    \right|^2 \right> = -\frac{4}{3} \epsilon^\pm l ,
\label{eq:theorem}
\end{equation}
where $\epsilon^\pm$ are the dissipation rates of the Els\"asser variables 
${\bf z}^\pm = {\bf u} \pm {\bf b}$, and the subindex $\parallel$ denotes 
the increment of the field along the displacement vector ${\bf l}$.
  
Moreover, even in the simplest incompressible case, at least 
two time scales can be identified in the inertial range of 
MHD turbulence. Besides the eddy turnover time, 
incompressible MHD flows are also characterized by the period of Alfv\'en 
waves $\tau \sim (B_0 L)^{-1}$, where $B_0$ is the amplitude of 
the large scale magnetic 
field in Alfvenic units. In a first attempt to derive a 
phenomenological theory, \citet{Iroshnikov63} and 
\citet{Kraichnan65} (IK) assumed that the large scale magnetic field acts 
as a uniform field for the small scale fluctuations, which then 
behave as Alfv\'en waves. In that case, small scales can interact not 
only through the eddies but also through Alfv\'en packages, which reduce 
the energy flux to small scales by increasing its transfer time. This 
introduces in practice a non-local interaction as the waves propagate 
along the large scale field (see \citealt{Gomez99} for a 
discussion). From dimensional analysis, Iroshnikov and Kraichnan then 
derived an isotropic energy spectrum proportional to $k^{-3/2}$. Later, 
extensions where considered to take into account the anisotropy induced 
at small scales by the large scale magnetic field 
\citep{Goldreich95,Galtier00,Galtier05,Boldyrev06}. Some of these extensions, 
after accounting for the anisotropy, rely on some form of a balance 
between the two fields that leaves only the turnover time as the 
relevant time scale, and can be therefore considered local or 
non-local depending on the authors.

At the core of the early disquisitions is the fact that in MHD 
the roles of a large scale flow and of a large scale magnetic field are 
different. While a (uniform) large scale flow can be removed by a 
Galilean transformation, a large scale magnetic field cannot. 
As a remarkable coincidence, lack of Galilean invariance is at the 
basis of the $\sim k^{-3/2}$ spectrum for hydrodynamic turbulence 
within the framework of the Direct Interaction Approximation (DIA) by 
\citet{Kraichnan59}, a flaw later corrected by the development of the 
Test Field Model (TFM) and the Lagrangian History DIA (LHDIA). However, 
in MHD magnetic fields are not Galilean invariant and for this 
reason the associated Alfv\'en waves have to be taken into account 
in phenomenological theories, and are also considered 
when studying non-local effects in the Eddy-Damped Quasi-Normal Markovian 
(EDQNM) closure \citep{Pouquet76} or in weak turbulence theory 
\citep{Nazarenko01}. However, although 
phenomenological descriptions assume that a large scale field has the 
effect of reducing the energy cascade rate, the transfer of energy (and 
the cascade) in many cases still takes place between eddies of similar 
size, presumably allowing for recovery of universal statistical 
properties at small scales.

In recent years, this universal behavior has been questioned by different 
authors. Because energy can be injected in MHD by a mechanical forcing 
or by an electromotive forcing, 
MHD turbulence is characterized by a larger number 
of regimes than hydrodynamic turbulence even in its simplest 
configurations. Magnetic fields in planets and stars are believed to be 
generated by dynamo action, where turbulent motions sustain magnetic fields 
against Ohmic dissipation \citep{Pouquet76,Moffatt,Krause,Brandenburg05}. 
This regime is often studied 
in numerical simulations (and recently in experiments, see 
\citealt{Monchaux07}) by mechanically stirring the flow. Depending on the 
amount of mechanical helicity in the flow (the alignment between the 
velocity and the vorticity), or in the presence of large-scale 
shear, the magnetic field generated may have large or small-scale 
correlation (compared with the integral scale of the flow), giving a steady 
state that may be dominated by mechanical or magnetic energy. On the other 
hand, plasmas in the solar corona and in the solar wind are dominated by 
magnetic energy, and are often studied numerically by stirring the flow 
with electromotive forces or using simulations of freely decaying 
turbulence. Finally, the amount of cross-correlation between the 
velocity and the magnetic field depends on the flow (e.g., on 
the heliocentric distance in the solar wind) and can also be varied 
in the simulations.

The questioning of universality was accompanied by recent detailed studies 
of scale interactions in MHD turbulence. Many of the 
studies considered the so called shell-to-shell transfer functions and 
partial energy fluxes, either in numerical simulations, observations, 
closures, or from the theoretical point of view. In the following 
sections we give a review of the results in this area, considering the 
several regimes studied by different authors, and also 
some examples of possible sources of non-locality in MHD. Finally, we 
discuss the results in the context of universality and of phenomenological 
theories for MHD. To briefly summarize the results, several authors have 
shown that the locality of energy transfer is in question in MHD flows. 
In particular, it was shown from simulations that the transfer of energy 
in MHD has two components: a local one that shares similar properties 
with hydrodynamic turbulence, and a component coupling the velocity and 
magnetic fields for which energy from the large scales can be under some 
circumstances injected directly into the small scales without the 
intervention of intermediate scales.

\section{Indirect evidence of non-locality}

Some theoretical, phenomenological, and (more recently) numerical results 
indicate scale interactions in MHD can be, under some conditions, 
of a different nature than in hydrodynamic turbulence. In this section 
we review early theoretical indications of non-locality in MHD turbulence, 
as well as numerical results that support the theoretical 
arguments without directly measuring scale interactions.

Early studies of dynamo action, and of magnetic field evolution under 
flows with simple strain, show that a large scale flow can excite, 
through field line stretching, magnetic fields at widely separated 
scales. One of the first works along these lines is the work of 
\cite{Batchelor50} where he considered the similarity between the 
induction equation and the vorticity equation 
($\vomega = \nabla \times {\bf u}$):
\begin{equation}
\frac{\partial \vomega}{\partial t} + {\bf u} \cdot \nabla \vomega = 
    \vomega \cdot \nabla {\bf u} + \nu \nabla^2 \omega .
\end{equation}
While the second term on the l.h.s.~advects the vorticity, the 
first term on the right (in three dimensions) produces vorticity by 
vortex stretching. For $P_M = \eta/\nu >1$ (the magnetic Prandtl number) 
Batchelor then concluded that the magnetic field would grow as magnetic 
field line stretching overcomes Ohmic dissipation. Later works 
considering stretching by uniform straining motion 
\citep{Moffatt64,Zeldovich84} showed that a large scale magnetic field 
can directly create small-scale magnetic fields. The work of 
\citet{Kazantsev68} considered a similar process under a random velocity 
field and described a non-local coupling that sustains the so-called 
small-scale dynamo, where magnetic fields are amplified at scales smaller 
than the integral scale of the flow. Several numerical simulations support 
these results, and show that smooth motions at the 
viscous scale give exponential growth of magnetic fields that 
can peak at the magnetic diffusion scale 
\citep{Schekochihin02,Schekochihin02b,Schekochihin04}.

The opposite limit, when the magnetic Prandtl number is much smaller 
than unity (a case of interest for industrial flows), is sometimes 
studied using the quasi-static approximation (see \citealt{Knaepen08} 
for a review). In this case, an external uniform magnetic field is 
applied, and the magnetic Reynolds number is chosen small enough that 
magnetic field fluctuations are rapidly damped. In that limit the 
Lorentz force in the momentum equation reduces to linear Joule damping 
\begin{equation}
\frac{\partial {\bf u}}{\partial t} + {\bf u}\cdot \nabla {\bf u} = 
    - \nabla p + \sigma B_0^2 \nabla^{-2} 
    \frac{\partial^2 {\bf u}}{\partial z^2} 
    + \nu \nabla^2 {\bf u} ,
\end{equation}
where $\sigma$ is the conductivity of the medium and the uniform magnetic 
field $B_0$ is in the $z$ direction. The Joule damping, although 
anisotropic in spectral space, is roughly independent of the wavenumber, 
and unlike viscous damping is not concentrated at small scales but rather 
acts at all scales. As a result, the large-scale magnetic field in this 
approximation exerts work over all scales in the velocity field (damping 
turbulent fluctuations) in a non-local way. We will see that the 
shell-to-shell transfers indicate in some cases similar behavior of the 
Lorentz force even in cases far from this approximation.

Another important example concerns Alfv\'en waves, which are also 
non-linear solutions of the ideal MHD equations. Alfv\'enic states with 
${\bf u} = \pm {\bf b}$ make the non-linear terms in Eqs. 
(\ref{eq:momentum}) and (\ref{eq:Induction}) zero, leaving only 
interactions with the large scale fields to transport energy across 
scales. Finally, it is worth mentioning here some recent attempts to 
build shell models of MHD turbulence (see e.g., 
\citealt{Plunian07,Stepanov08,Lessinnes09}). In these models, it was 
found that many features of steady state MHD turbulence 
can be reproduced using local coupling between shells, 
but that to reproduce the small-scale dynamo and turbulence 
at $P_M \gg 1$, non-local transfers have to be considered \citep{Stepanov08}.

\section{The shell-to-shell transfer}

In recent years, the increase in computing power allowed numerical 
exploration of MHD turbulence in different regimes. The development of 
shell-to-shell transfers (see \citealt{Dar01,Debliquy05,Alexakis05a}) 
allowed for explicit computation of detailed scale interactions in MHD 
turbulence using the output stemming from the simulations and 
without the need to compute the more expensive triadic interactions. In 
this section we briefly introduce the isotropic shell-to-shell energy 
transfer functions, and describe how fluxes can be obtained from them.

A shell filter decomposition of the two fields is introduced as
\begin{equation}
{\bf u}({\bf x}) = \sum_K {\bf u}_K({\bf x}) , 
\end{equation}
\begin{equation}
{\bf b}({\bf x}) = \sum_K {\bf b}_K({\bf x}) ,
\end{equation}
where
\begin{equation}
{\bf u}_K({\bf x})=\sum_{K_1<|{\bf k}|\le K_2}{\bf \tilde{u}}({\bf k})
    e^{i{\bf k}\cdot{\bf x}} 
\end{equation}
and
\begin{equation}
{\bf b}_K({\bf x})=\sum_{K_1<|{\bf k}|\le K_2}{\bf \tilde{b}}({\bf k})
    e^{i{\bf k}\cdot{\bf x}} .
\end{equation}
Here $\tilde{\bf u}({\bf k})$ and $\tilde{\bf b}({\bf k})$ are respectively 
the Fourier transforms of the velocity and magnetic fields with wavenumber 
${\bf k}$. The shell filtered fields ${\bf u}_K$ and ${\bf b}_K$ 
are therefore defined as the field 
components whose Fourier transforms contain only 
wavenumbers in a given shell $K$. These shells can be defined with linear 
binning using $K_1=K$ and $K_2=K+1$, or alternatively with logarithmic 
binning using $K_1=\gamma^n K_0$ and $K_2=\gamma^{n+1} K_0$ 
for some positive $\gamma>1$ and integer $n$ ($\gamma=2$ is often used). 
The latter definition has the 
advantage of being conceptually closer to the idea of ``scale'' of 
eddies in turbulence, which in general implies logarithmic division 
of wavenumbers. The former has the advantage of having a direct 
association with Alfv\'en waves, which are of the form 
${\bf u} = \pm {\bf b} \sim e^{i({\bf k}\cdot{\bf x} \pm \omega t)}$ in 
periodic boxes or in infinite domains, and which are more akin to linear 
treatment of spectral space. Note that the transfer among logarithmic 
shells can be reconstructed by summing over the linearly spaced shells.

Another variant when defining the shell filter decomposition has to do with 
the choice of using sharp filters (as in the equations above) or smooth 
filters \citep{Eyink94,Eyink05}. This issue has raised some controversy in 
the hydrodynamic case, with claims that non-localities observed in 
simulations may be due to the commonly used sharp filters. 
Recent numerical comparisons 
\citep{Domaradzki07a,Domaradzki07b} have shown that results are only weakly 
dependent on the shape of the filter used, except in the case where a very 
broad smooth filter is considered. Moreover, recent theoretical results 
were able to show locality of hydrodynamic turbulence in Fourier space 
in the limit of infinite Reynolds numbers for both smooth and sharp filters 
\citep{Eyink09,Aluie09}.

Based on the shell filter decomposition, the evolution of the kinetic 
energy in a shell $K$, $E_u(K)=\int {\bf u}_K^2/2 \, dx^3$, can be derived 
from Eq. (\ref{eq:momentum}) as
\begin{equation}
\frac{\partial E_u(K)}{\partial t} = 
    \sum_Q \left[T_{uu}(Q,K)+T_{bu}(Q,K)\right] - \nu D_u(K) ,
\label{eq:Eu}
\end{equation}
and for the magnetic energy, $E_b(K)=\int {\bf b}_K^2/2 \, dx^3$, from Eq. 
(\ref{eq:Induction})
\begin{equation}
\frac{\partial E_b(K)}{\partial t} =
    \sum_Q \left[T_{bb}(Q,K)+{T}_{ub}(Q,K)\right] - \eta D_b(K) ,
\label{eq:Eb}
\end{equation}
where the functions $D_u(K)$ and $D_b(K)$ express respectively the 
kinetic and magnetic energy dissipation in the shell $K$. The transfer 
functions $T_{uu}(Q,K)$, ${T}_{ub}(Q,K)$, ${T}_{bb}(Q,K)$, and 
${T}_{bu}(Q,K)$, that express the energy transfer between different fields 
and shells are given by
\begin{eqnarray}
T_{uu}(Q,K) & = &
-\int {{\bf u}_K \left({\bf u} \cdot \nabla\right) {\bf u}_Q  } dx^3, \\ 
T_{bu}(Q,K) & = &
\int {{\bf u}_K ({\bf b}\cdot \nabla) {\bf b}_Q  } dx^3, \\
T_{bb}(Q,K) & = &
-\int {{\bf b}_K ({\bf u} \cdot \nabla) {\bf b}_Q  } dx^3, \\
T_{ub}(Q,K) & = &
\int {{\bf b}_K ({\bf b} \cdot \nabla) {\bf u}_Q  } dx^3.
\end{eqnarray}
The function $T_{uu}(Q,K)$ measures the transfer rate of kinetic energy 
in the shell $Q$ to kinetic energy in the shell $K$ due to the 
advection term in the momentum equation (\ref{eq:momentum}). This is 
the non-linear transfer that is also present in hydrodynamic turbulence. 
Similarly, $T_{bb}(Q,K)$ expresses the rate of magnetic energy transferred 
from the shell $Q$ to magnetic energy in the shell $K$ 
due to the magnetic advection term. The Lorentz force is responsible for 
the transfer of energy from the magnetic field in the shell $Q$ to the 
velocity field in the shell $K$, as measured by $T_{bu}(Q,K)$. Finally, 
the term responsible for the stretching of magnetic field lines, the 
first term on the r.h.s.~of Eq. (\ref{eq:Induction}), results in the 
transfer of kinetic energy from shell $Q$ to magnetic energy in shell $K$ 
and is expressed by $T_{ub}(Q,K)$. This is the term that describes 
magnetic induction and dynamo action. 

All these transfer functions satisfy
\begin{equation}
\label{tran_id}
T_{vw}(Q,K)=-T_{wv}(K,Q) .
\end{equation}
(where the subindices $v$ and $w$ stand for $u$ or $b$). 
This expression indicates
that the rate at which the shell $Q$ gives energy to the shell
$K$ is equal to the rate the shell $K$ receives energy from
the shell $Q$, and is a necessary condition to define shell-to-shell 
transfers that satisfy a detailed energy balance between shells. 
Then, the contribution of these transfers to the total energy flux can 
be computed as:
\begin{equation}
\Pi_{vw}(k) = -\sum_{K=0}^k \sum_Q T_{vw}(K,Q) .
\end{equation}

Besides the total energy, the MHD equations have two more ideal 
invariants: the cross-helicity 
$C=\int {\bf u}\cdot{\bf b} \, dx^3$, and the magnetic 
helicity $H=\int {\bf a}\cdot{\bf b} \, dx^3$ where ${\bf a}$ is the 
vector potential such as ${\bf b}=\nabla \times {\bf a}$. These quantities 
also satisfy detailed balance equations equivalent to Eqs. (\ref{eq:Eu}) and 
(\ref{eq:Eb}). Shell-to-shell transfer functions for the magnetic helicity 
have been defined in \citet{Alexakis06}. Its transfer from shell $Q$ to 
shell $K$ is measured by
\begin{equation}
T_H(K,Q) = \int{ {\bf b}_K \cdot \left( {\bf u}_K \times {\bf b}_Q \right) 
dx^3 }.
\end{equation}

The energy transfer functions were also generalized in recent works to 
consider the flux of energy in terms of the Els\"asser variables 
\citep{Alexakis05a,Carati06,Alexakis07}, anisotropic transfers 
\citep{Alexakis07,Teaca09}, forward and backward transfers in an attempt to 
quantify backscatter required for subgrid models 
\citep{Debliquy05,Carati06}, and extensions to consider 
compressibility effects \cite{Pietarila10}, and kinetic plasma effects 
as in two-fluid MHD approximations \citep{Mininni07}.

\section{Direct studies of multi-scale interactions}

The shell-to-shell energy transfers have been studied
extensively \citep{Dar01,Verma04,Alexakis05a,Mininni05b,Debliquy05,Carati06} 
for a variety of mechanically forced and decaying MHD flows in two and three 
dimensions. Depending 
on the configuration, different degrees of non-locality were reported. In 
the following subsections we present a short summary of the results 
discriminating by the forcing configuration. Overall, we can say that 
in all cases examined in the literature the transfers $T_{uu}$ and $T_{bb}$ 
have a local behavior: energy is transferred forward between nearby shells, 
in a fashion similar to what is observed in hydrodynamic turbulence 
\citep{Domaradzki90,Ohkitani92,Zhou93,Yeung95,Alexakis05b,Mininni06}. On 
the other hand, the $T_{bu}$ and $T_{ub}$ transfers that express the 
energy exchange between the velocity and the magnetic field have a rather 
different behavior.

\subsection{Forced isotropic and homogeneous turbulence}

\begin{figure}
\centerline{
\epsfxsize=7.5cm
\epsfysize=5cm
\epsfbox{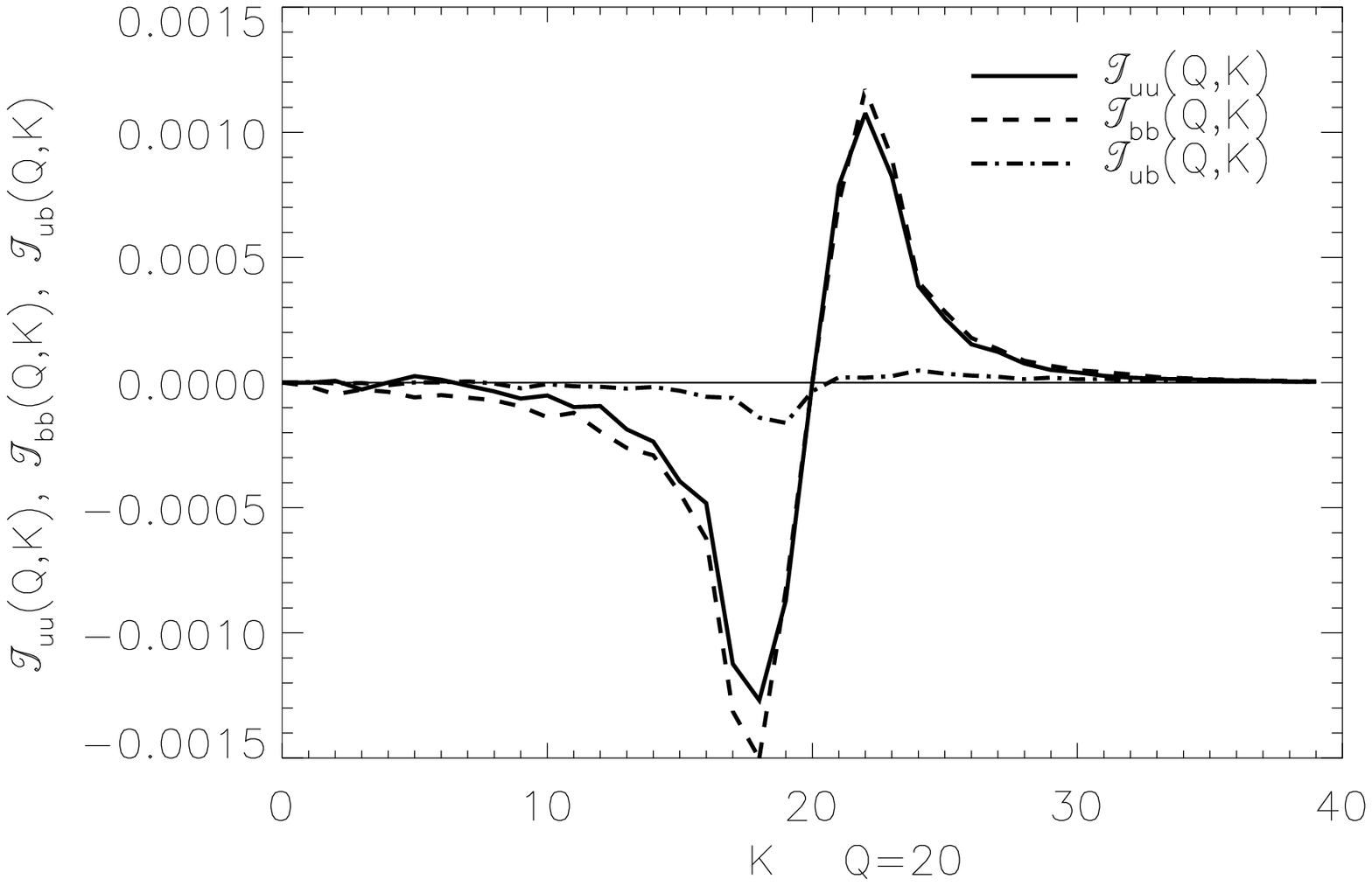} 
\epsfxsize=6.5cm
\epsfysize=5cm
\epsfbox{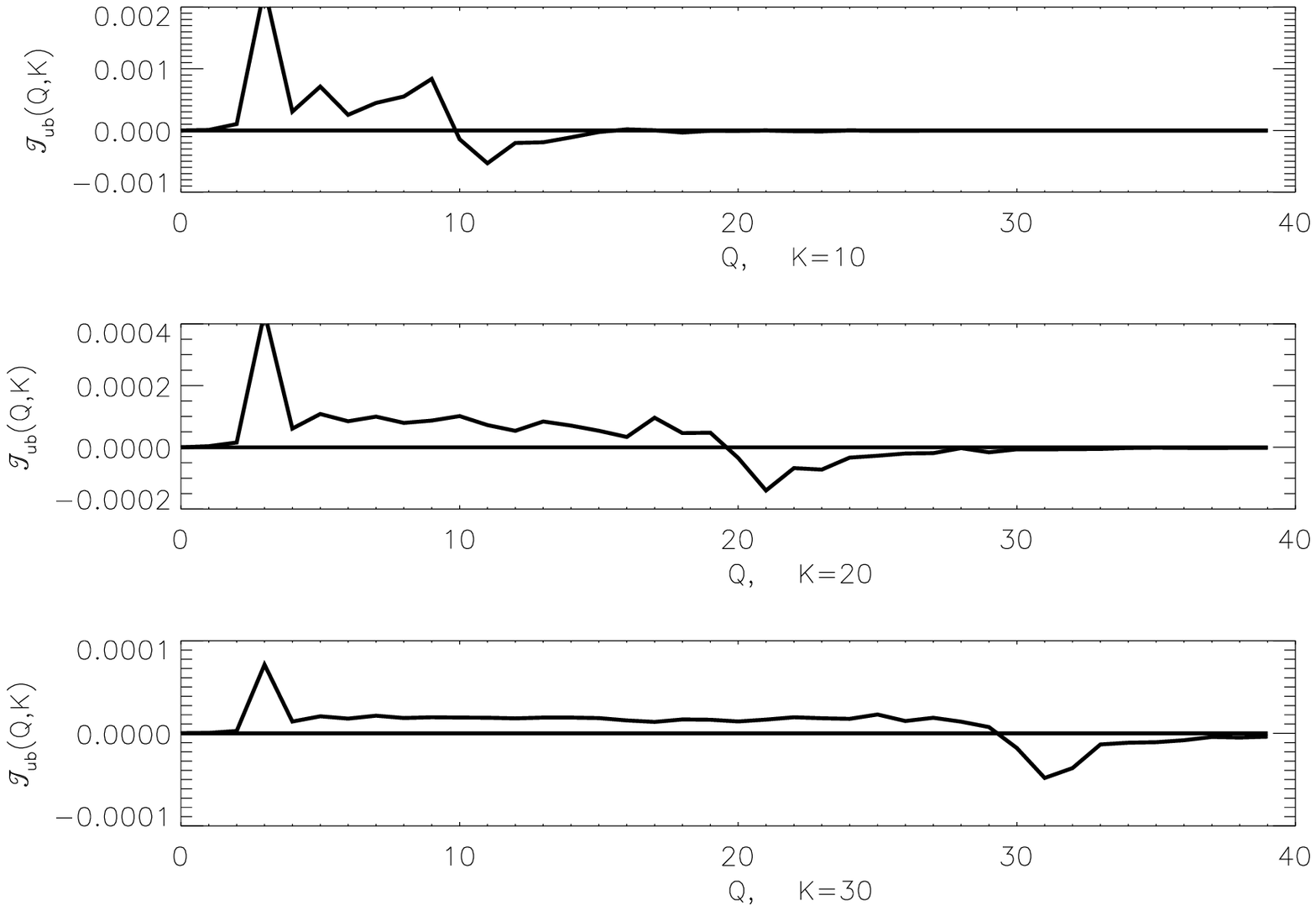}}
\caption{Left: transfer functions in mechanically forced MHD turbulence, 
for $Q=20$. The $T_{uu}$ and $T_{bb}$ functions are local, with a negative 
peak for $K<Q$ and a positive peak for $K>Q$, which indicate energy is 
removed by these transfer functions from smaller wavenumbers and given 
to slightly larger wavenumbers. The transfer between magnetic to kinetic 
energy is of smaller amplitude and also seems local. Right: The $T_{ub}$ 
transfer, for different values of $K$. This function is non-local, with 
a strong peak at the forcing scale and with a constant positive plateau 
that extends up to $K \approx Q$. Adapted from \citet{Alexakis05a}.}
\label{fig1}
\end{figure}

As mentioned before, forced simulations of MHD turbulence can be attained 
by forcing both fields, or by forcing only the velocity field (in which 
case magnetic fields are sustained by dynamo action; a distinction must 
be done then between the kinematic regime, where the magnetic field has no 
backreaction on the flow, and the turbulent steady state where the magnetic 
field modifies the flow through the Lorentz force). The mechanically forced 
case is of more interest as it is closer to astrophysical and geophysical 
configurations, and as it is consistent with the constraint of magnetic 
flux conservation. The first studies of shell-to-shell transfer from 
simulations in such configuration were presented in \citet{Alexakis05a}. 
In the simulations with a resolution of $256^3$ grid points, the velocity 
field was forced with time independent mechanical forcing until a 
hydrodynamic turbulent steady state was reached. Two different forcing 
functions were studied: one non-helical and one helical. Magnetic Prandtl 
numbers of unity and smaller were considered. Once a hydrodynamic steady 
state was reached for each forcing function, a small magnetic field was 
introduced and, after the transient kinematic dynamo amplification, the 
system reached a steady state MHD turbulent regime. In such a state the 
transfer functions described in the previous section were computed. Typical 
results are illustrated in Fig. \ref{fig1}.

The $T_{uu}$ and $T_{bb}$ transfers were observed to behave in a similar 
fashion, giving direct and local transfer of energy. In Fig. \ref{fig1}, 
this is indicated by the negative and positive peaks, which show 
energy is removed by these transfer functions from smaller wavenumbers and 
given to slightly larger wavenumbers. However, for $T_{ub}$ a distinct 
behavior appeared. The large scale flow injected energy (through 
stretching) directly into the magnetic field at all scales. This manifests 
itself as a peak at the mechanical forcing scale for all receiving shells, 
and as an extended positive plateau (note positive $T_{ub}$ 
indicates energy given by the velocity field at shell 
$Q$ to magnetic field at shell $K$). 
In other words, at a given shell $K$, the magnetic field receives energy 
from the velocity field in all shells $Q<K$, and gives energy to the 
velocity field in shells $Q>K$. This result, reminiscent of the theoretical 
arguments by \citet{Batchelor50} and \citet{Zeldovich84}, was interpreted 
as the sustainment of the magnetic field against Ohmic dissipation by 
dynamo action: to maintain the magnetic field when only the 
velocity field is stirred, a nonzero flux from the velocity field 
to the magnetic field 
is required at all times. It is worth pointing out here that in the steady 
state this non-local transfer is small compared with the local transfers 
(approximately $10-20\%$ at the resolutions studied). When considering 
Els\"asser variables, the transfer functions were observed to become 
more local.

The case of random forcing with magnetic Prandtl number of unity was 
studied in \citet{Carati06} using $512^3$ simulations. The analysis, which 
used logarithmic binning, confirmed the previous results, showing local 
transfer in $T_{uu}$ and $T_{bb}$, and non-local coupling between 
the velocity and the magnetic field. This indicates that the phenomenon 
may be independent of the type of forcing, and associated to the stretching 
process that sustains the magnetic field. The work also discussed the 
possibility of splitting the transfer functions to discriminate between 
forward and backward contributions, which were used to discuss 
implications of the shell-to-shell transfers for LES models. 
Similar results were obtained for forced two dimensional 
MHD turbulence \citep{Dar01}.

A different approach was considered by \citet{Yousef07}, who studied 
the steady state of small-scale dynamo action for $P_M \le 1$. Instead 
of using transfer functions, to measure the different components 
of the energy flux they considered the Politano-Pouquet relations 
(\ref{eq:theorem}) in terms of the velocity and the magnetic field
\begin{equation}
\left< \delta u_\parallel \left( \left|\delta {\bf u}\right|^2 + 
    \left|\delta {\bf b}\right|^2 \right) \right> \mp 
\left< \delta b_\parallel \left( \left|\delta {\bf u}\right|^2 + 
    \left|\delta {\bf b}\right|^2 \right) \right> \pm 
2 \left< \delta {\bf u} \cdot \delta {\bf b} \left( \delta u_\parallel 
    \mp \delta b_\parallel \right) \right> = -\frac{4}{3} \epsilon^{\pm} l ,
\end{equation}
together with \citet{Chandrasekhar51} law
\begin{equation}
\left< \delta u_\parallel^3 \right> -6 \left< b_\parallel^2 \delta 
    u_\parallel \right> = -\frac{4}{5} \epsilon l ,
\end{equation}
where $\epsilon$ is the total energy flux. The authors 
discriminated between the 
different terms to look how they balanced to give rise to the 
direct flux. Each of the terms in these expressions can indeed 
be associated to the counterpart in real space of the 
$\Pi_{uu}$, $\Pi_{bb}$, and $\Pi_{ub}+ \Pi_{bu}$ fluxes 
in Fourier space.

The dominant balance was identified between $(4/5) \epsilon l$ and 
$6 \left< b_\parallel^2 \delta u_\parallel \right>$, and they concluded 
that at their available resolution, the local direct cascade of energy 
was ``short-circuited'' by the transfer of kinetic energy into magnetic 
energy. They also associated this non-local coupling with the folded 
structure of the small-scale magnetic field. Using the shell-to-shell 
transfer approach, \citet{Alexakis07} further showed that the non-local 
effects disappear if phases are randomized for the two fields, which also 
make the current sheet and folded structures disappear.

The non-local effects play a more important role in the kinematic dynamo 
regime \citep{Mininni05b}, as in that case the turbulence is not in a 
steady regime and $T_{ub}$ accounts for all mechanisms that amplify 
the magnetic field. In that case, the $T_{ub}$ transfer has been shown to be
useful to identify and quantify scale-by-scale sources of dynamo action 
\citep{Mininni05b,Alexakis07b}.

\subsection{Freely decaying turbulence}

\begin{figure}
\centerline{
\epsfxsize=14cm
\epsfbox{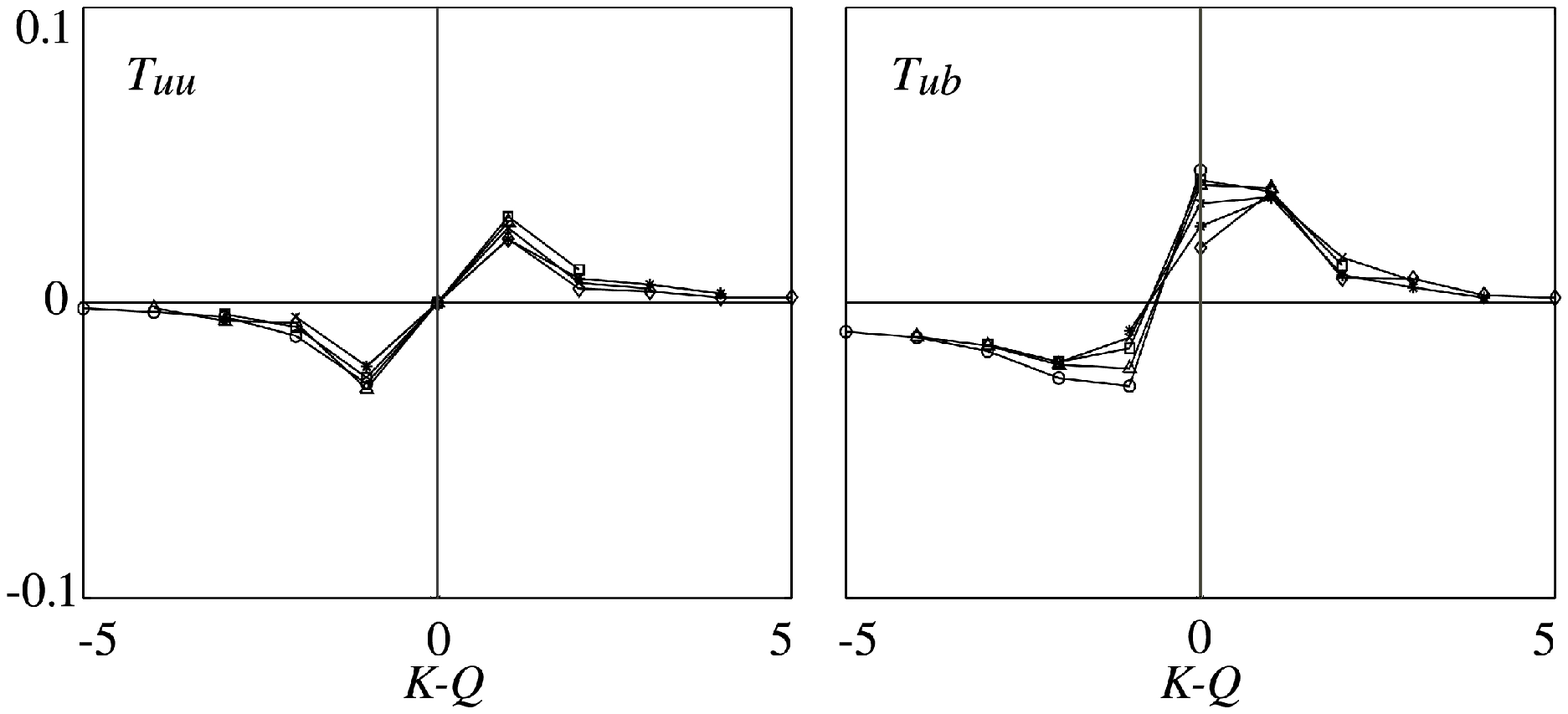}}
\caption{Left: $T_{uu}$ transfer function in freely decaying MHD 
turbulence, for different shells. The $T_{bb}$ transfer function is 
similar but has twice the amplitude. Right: $T_{bu}$ transfer function 
in the same simulation. Note the peak for $K-Q=0$, indicating most 
interchange of energy between the velocity and the magnetic field takes 
place between similar scales. The shells are logarithmically binned. 
Adapted from \citet{Debliquy05}.}
\label{fig2}
\end{figure}

The non-local effects observed in forced turbulence are either absent 
or negligible in the freely decaying case. In \citet{Debliquy05}, $512^3$ 
simulations of freely decaying MHD turbulence were considered. The 
$T_{uu}$ and $T_{bb}$ transfers are similar to the forced case (see Fig. 
\ref{fig2}) and indicate local direct transfer. However, the $T_{ub}$ and 
$T_{bu}$ transfer functions were also observed to be local, with most of 
the transfer between the velocity and the magnetic field taking place 
between the same shell. The remaining transfer (for non-neighboring shells) 
was observed to decay more slowly than in the $T_{uu}$ and $T_{bb}$ functions, 
but except for this detail no other indications of non-locality were 
reported.

Similar results were obtained from analysis of solar wind turbulence 
\citep{Strumik08a,Strumik08b}. Solar wind turbulence is often considered 
the MHD equivalent of hydrodynamic freely decaying wind tunnel turbulence 
(see \citealt{Bruno05} for a review). From 1996 Ulysses magnetometers 
time series and using a Markov process approach, 
\citet{Strumik08a} concluded that the transfer of magnetic to magnetic 
energy was local. Then, using velocity and magnetic field time series 
from ACE spacecraft from 1999 to 2006 and performing the same analysis 
on the remaining transfers they concluded that all transfers were local.

The differences between the forced and freely decaying cases can be 
understood as in the mechanically forced runs, the velocity field has 
to continuously supply energy to the magnetic field in order to sustain 
it against Ohmic dissipation. This is not necessarily the case for 
freely decaying runs where both fields are dissipated in time.

\subsection{Anisotropic turbulence}

\begin{figure}
\centerline{
\epsfxsize=14cm
\epsfbox{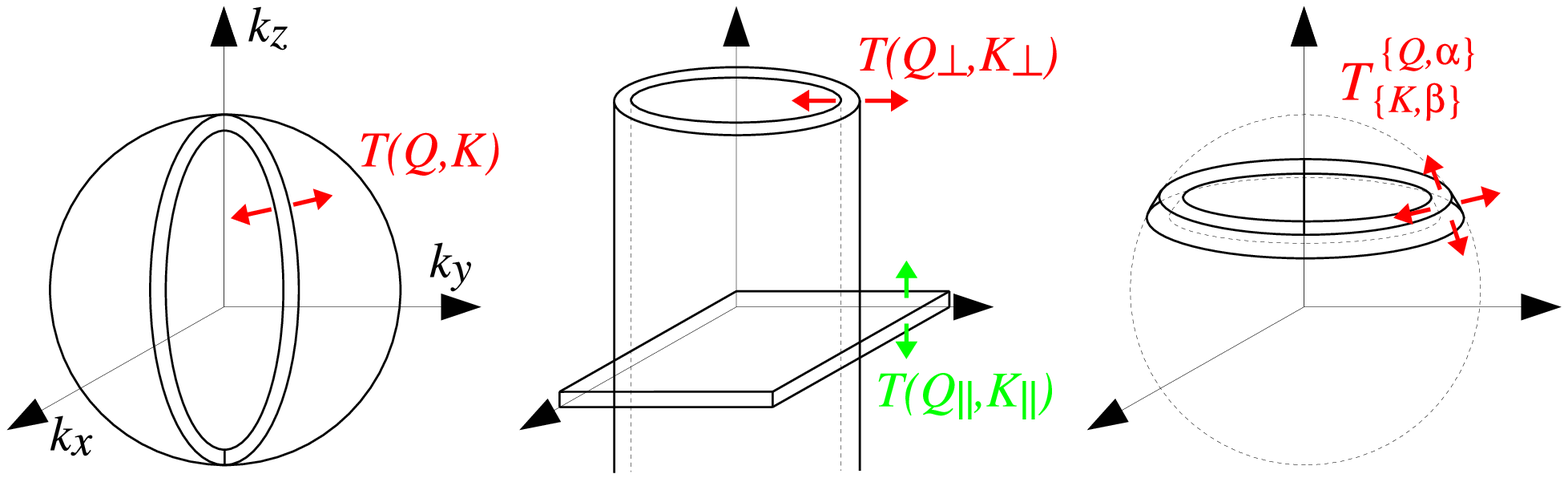}}
\caption{Isotropic (spherical) shells in the left, and anisotropic 
foldings of shells in Fourier space. The uniform magnetic field is 
assumed to be in the $z$ direction. Cylindrical and planar shells are 
shown in the middle, and ring shells are shown in the right. Transfer 
of energy across planes is denoted by $T(Q_\parallel,K_\parallel)$, 
and transfer across cylinders is denoted by $T(Q_\perp,K_\perp)$. For 
ring-to-ring transfers, the notation $T^{\{Q,\alpha\}}_{\{K,\beta\}}$ 
denotes transfer can be measured between $K$ and $Q$ spherical shells, 
as well as between two azimuthal angles $\alpha$ and $\beta$.}
\label{fig3}
\end{figure}

Recently, the shell-to-shell transfers were extended to consider 
anisotropies when an external uniform magnetic field is imposed. This 
case is of interest as in many astrophysical problems a strong 
large-scale magnetic field is present creating small-scale anisotropy. 
Unlike hydrodynamic turbulence, MHD turbulence does not recover 
isotropy at small scales, and theoretical and numerical results 
indicate anisotropy becomes stronger at smaller scales.

\begin{figure}
\centerline{
\epsfxsize=6.5cm
\epsfysize=7.5cm
\epsfbox{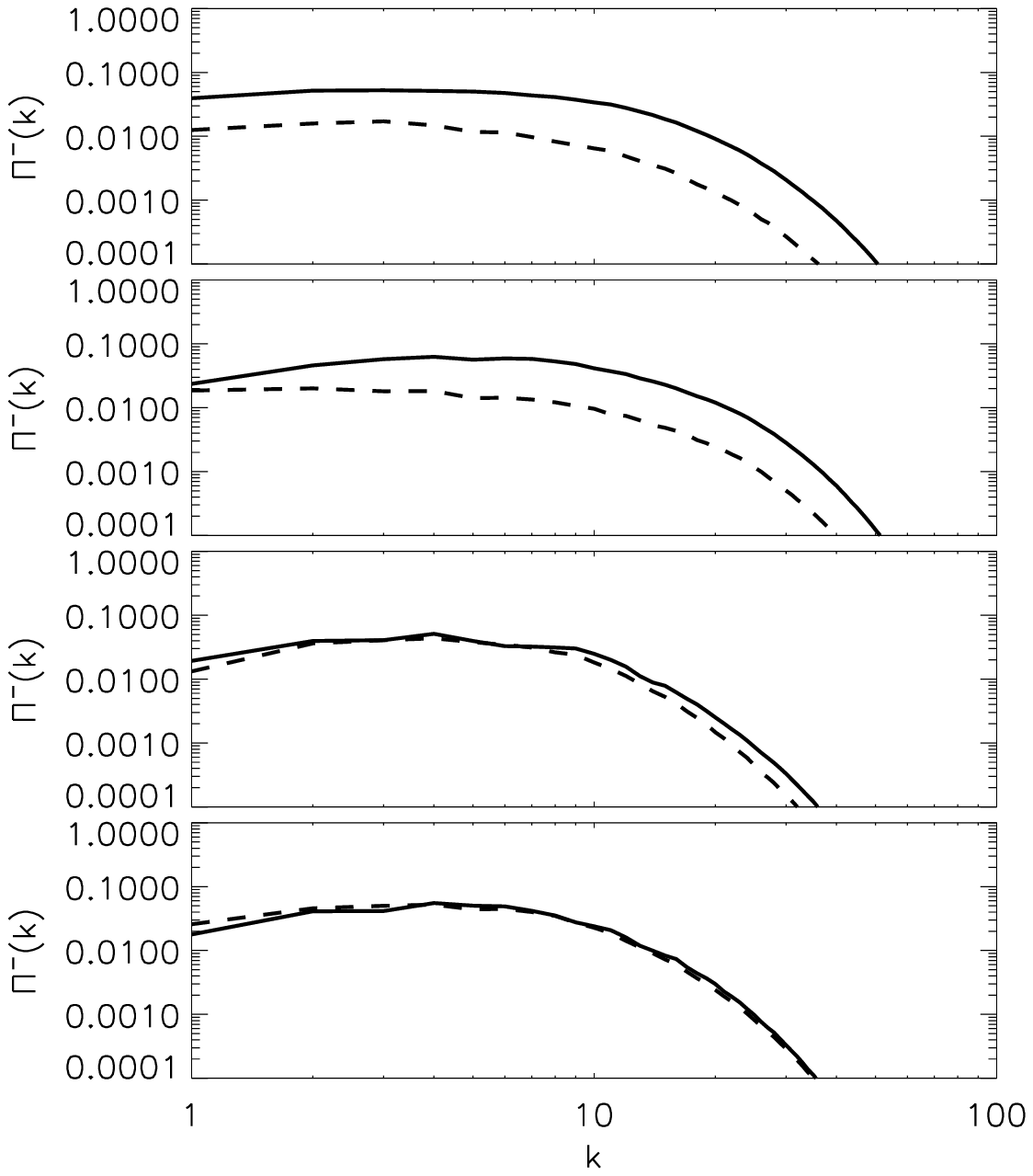} 
\epsfxsize=6.5cm
\epsfysize=7.5cm
\epsfbox{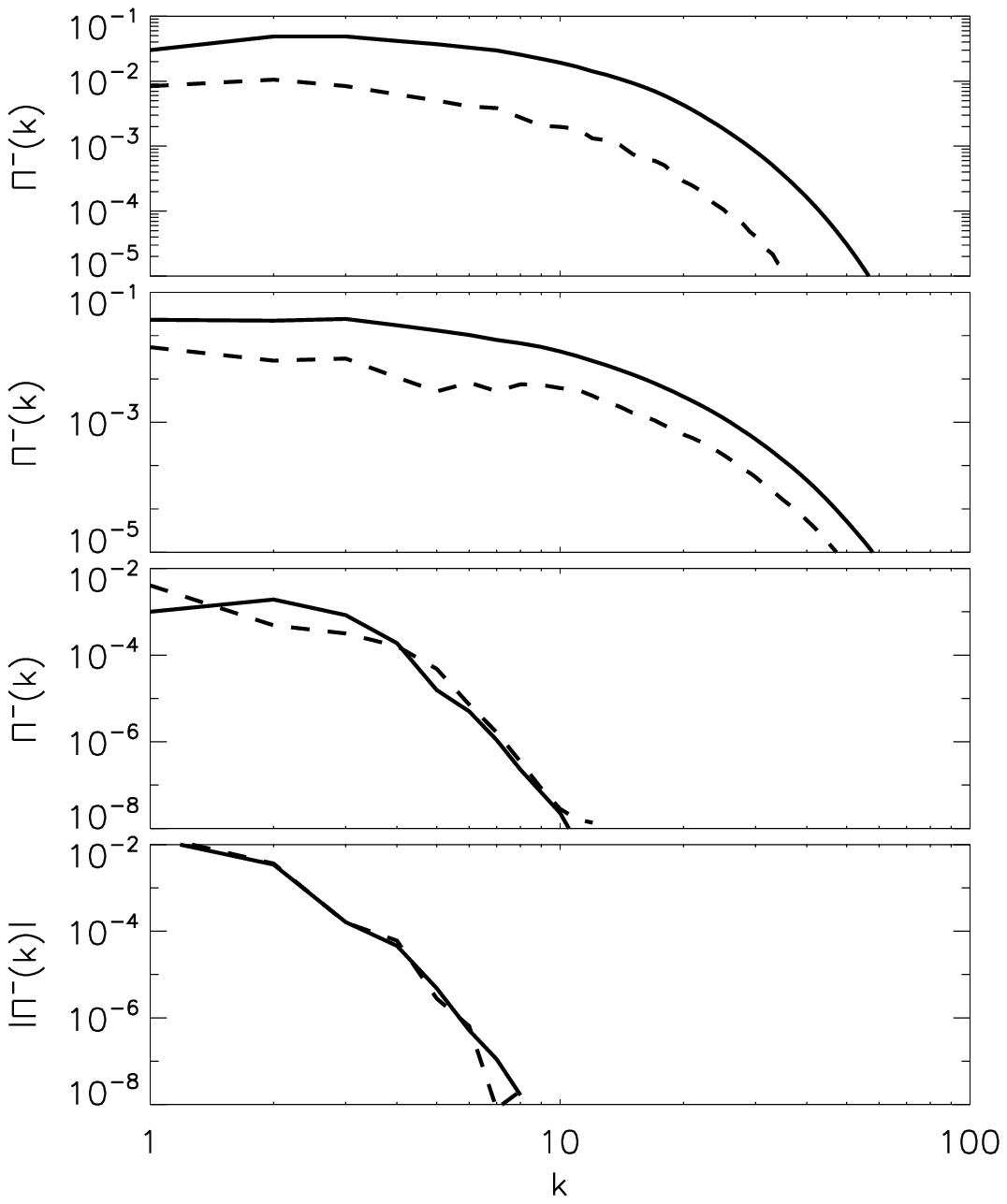}}
\caption{Left: total energy flux (solid) across cylinders and partial 
flux associated to interactions with modes with $k_\parallel=0$ (dashed), 
with four different values of the external magnetic field $B_0$ from 
0 to 15 (from top to bottom). Right: same but with the total flux 
and partial flux associated to interactions with modes with 
$k_\parallel=1$ across planes. From \citet{Alexakis07}.}
\label{fig4}
\end{figure}

To study anisotropic transfers, different foldings of the shells in 
Fourier space can be implemented. Fig \ref{fig3} shows the possible 
options. In \citet{Alexakis07}, anisotropic shell-to-shell transfer 
functions were introduced folding Fourier shells in cylinders (associated 
to wavenumbers $k_\perp$ perpendicular to the mean magnetic field) and in 
planes (associated to parallel wavenumbers $k_\parallel$). Shell-to-shell 
transfers were only considered for the Els\"asser variables, but the 
fluxes were reconstructed from these functions to measure the relative 
contribution of non-locality to the total flux. Freely decaying simulations 
with spatial resolution of $256^3$ grid points were analyzed, and the 
amplitude of the imposed magnetic field was varied from 0 to $15$ (in 
units of the initial small scale fluctuations). The transfer functions of 
the two Els\"asser energies were found local in both parallel and 
perpendicular directions, irrespectively of the amplitude of the external 
field. However, interactions between the counterpropagating Alfv\'en 
waves were reported to become non-local. For strong magnetic fields, most 
of the energy flux in the perpendicular direction was found to be due to 
interactions with modes with $k_\parallel = 0$ (see Fig. \ref{fig4}). 
In the parallel direction, however, $k_\parallel = 0$ modes cannot 
transfer energy and most of the interactions were 
observed to take place with modes near 
$k_\parallel \approx 0$. The results are in qualitative agreement with 
predictions from weak turbulence theory \citep{Galtier00} and with recent 
non-local phenomenological models \citep{Alexakis07c}.

A different approach to study anisotropic transfers was presented by 
\citet{Teaca09}, who decomposed the spectral space into rings, studying 
then transfers along radial and angular directions in spectral space 
(which they termed ``ring-to-ring'' transfers). They considered 
forced simulations of MHD turbulence with an 
imposed magnetic field, with a spatial resolution of $512^3$ grid points 
and varying the imposed magnetic field between 0 to $\sqrt{10}$ (in 
units of the small-scale magnetic field fluctuations). They also observed 
dominance of energy transfer in the direction perpendicular to the 
uniform magnetic field, and suppression of the transfer in the parallel 
direction. Their approach is useful to understand how energy is 
angularly distributed in spectral space to create anisotropy. 
Non-local effects with the forcing shell were observed in the 
shell-to-shell transfers, but in the angular ring-to-ring transfers were 
too weak to be noticed.

\subsection{Magnetic helicity and the inverse cascade}

\begin{figure}
\centerline{
\epsfxsize=14cm
\epsfbox{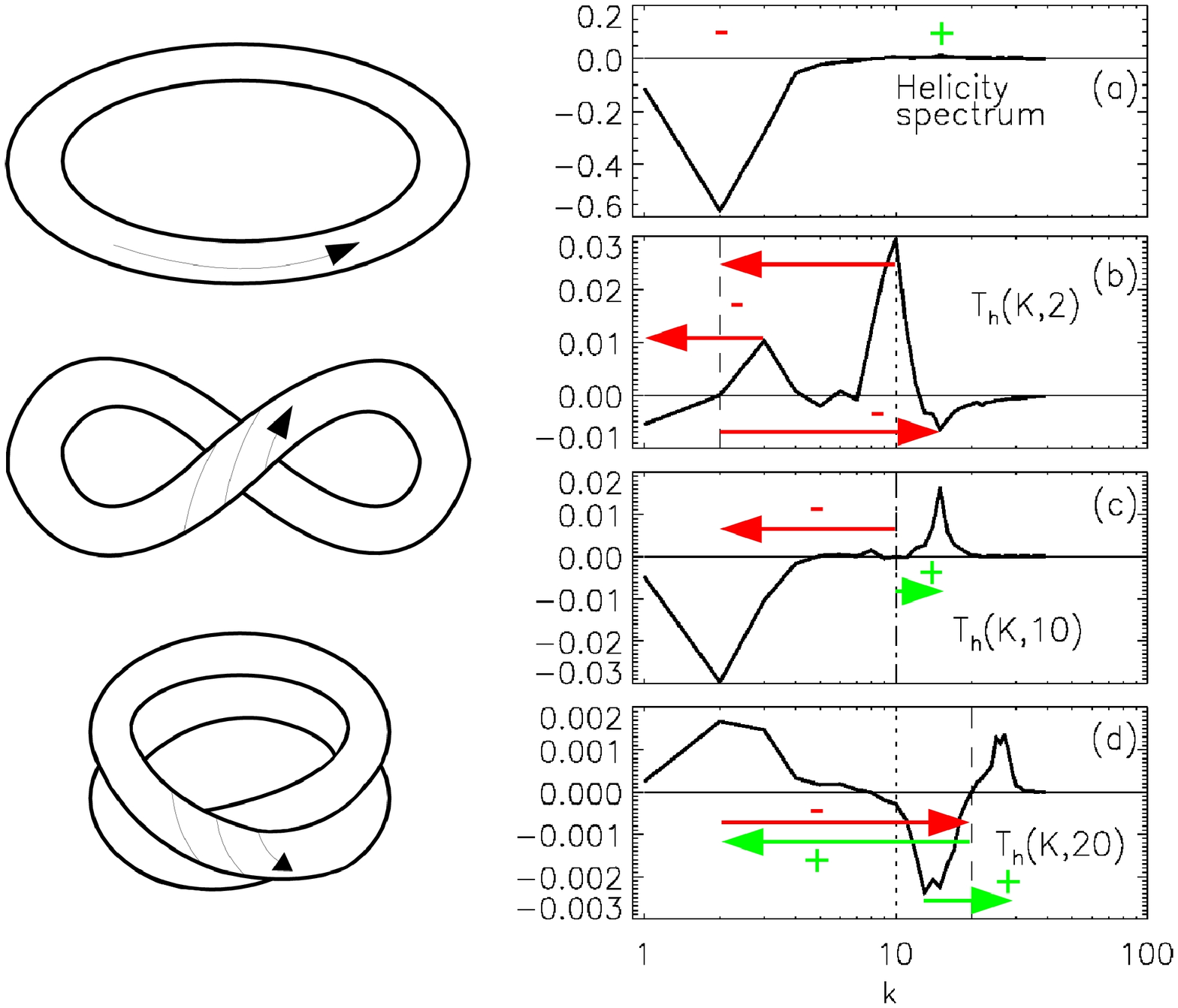}}
\caption{Left: the stretch, twist, and fold dynamo mechanism. Each time 
a closed magnetic flux tube is twisted, magnetic helicity of opposite sign 
is created at large and small scales. The folding creates regions where 
helical magnetic fields can reconnect. Right: 
(a) The helicity spectrum in a simulation with (positive) helical 
mechanical forcing at $k=10$. Magnetic helicity is negative at larger scales 
and positive at smaller scales. (b,c,d): The transfer of helicity 
for $Q=2$, 10, and 20. The red arrows indicate transfer of negative 
helicity and the green arrows transfer of positive helicity. At large 
scales (b), negative magnetic helicity inversely cascades locally between 
neighbor shells and non-locally from the forced shell and to the small scale 
shells. At the forced shell (c), the forcing injects opposite signs 
of helicity at large and small scales. At small scales (d), positive 
magnetic helicity has a local direct transfer of helicity, while the 
small scales also remove negative magnetic helicity from the large scales. 
Note direct transfer of negative helicity is equivalent to inverse transfer 
of positive helicity.}
\label{fig5}
\end{figure}

Non-local transfers were also reported in investigations of the cascade of 
magnetic helicity. Magnetic helicity is an ideal invariant in MHD, that is 
known to cascade inversely (to the large scales) in a turbulent flow
\citep{Pouquet76,Meneguzzi81,Brandenburg01,Gomez04,Brandenburg05,Alexakis06}. 
The generation of large scale magnetic fields in 
galaxies and other astrophysical bodies is sometimes attributed to 
this inverse cascade. In \citet{Alexakis06} magnetically and mechanically 
forced simulations were considered. In both cases, both local and 
non-local transfers were observed. At early times, magnetic helicity was 
observed to cascade inversely and locally from the closest neighbor shells, 
and non-locally from the forced shells. When the correlation length became 
of the size of the box, the direct input from the forced scales became 
dominant, and a local direct transfer of helicity from large to small 
scales also developed. This latter effect was speculated to be 
dependent on boundary conditions and therefore non-universal.

In the mechanically forced case the inverse cascade of 
helicity was associated to the large-scale dynamo $\alpha$-effect 
\citep{Steenbeck66,Pouquet76,Krause,Brandenburg01,Brandenburg05}. 
In that case, the mechanical forcing creates equal amounts of magnetic 
helicity of opposite signs at large and small scales. The process can be
understood using the conceptual ``stretch, twist, and fold'' (STF) 
dynamo mechanism \citep{Vainshtein72,Childress}. Each time a closed 
magnetic flux tube is twisted by the helical velocity field, 
magnetic helicity is created at large scales, while small scale 
magnetic field lines are twisted in the opposite direction thus 
creating equal amount of magnetic helicity of opposite sign in the 
small scales. As the STF process is repeated, the large-scale helicity 
is transferred inversely both locally and non-locally (with constant 
flux), while the small-scale helicity is pushed towards smaller scales 
(see Fig \ref{fig5}). This latter process removes magnetic 
helicity from the large scales and allows the magnetic 
field to ``disentangle'' through reconnection events, destroying in that 
way magnetic helicity \citep{Alexakis06,Alexakis07b}. At this moment it 
is unclear whether this processes should be associated to a 
cascade (i.e., if the process takes place with constant flux), although 
results in \citet{Alexakis07b} and \citet{Mininni09} suggest this may 
not be the case.

\section{Non-local interactions and universality of MHD turbulence}

\begin{figure}
\centerline{
\epsfxsize=9cm
\epsfbox{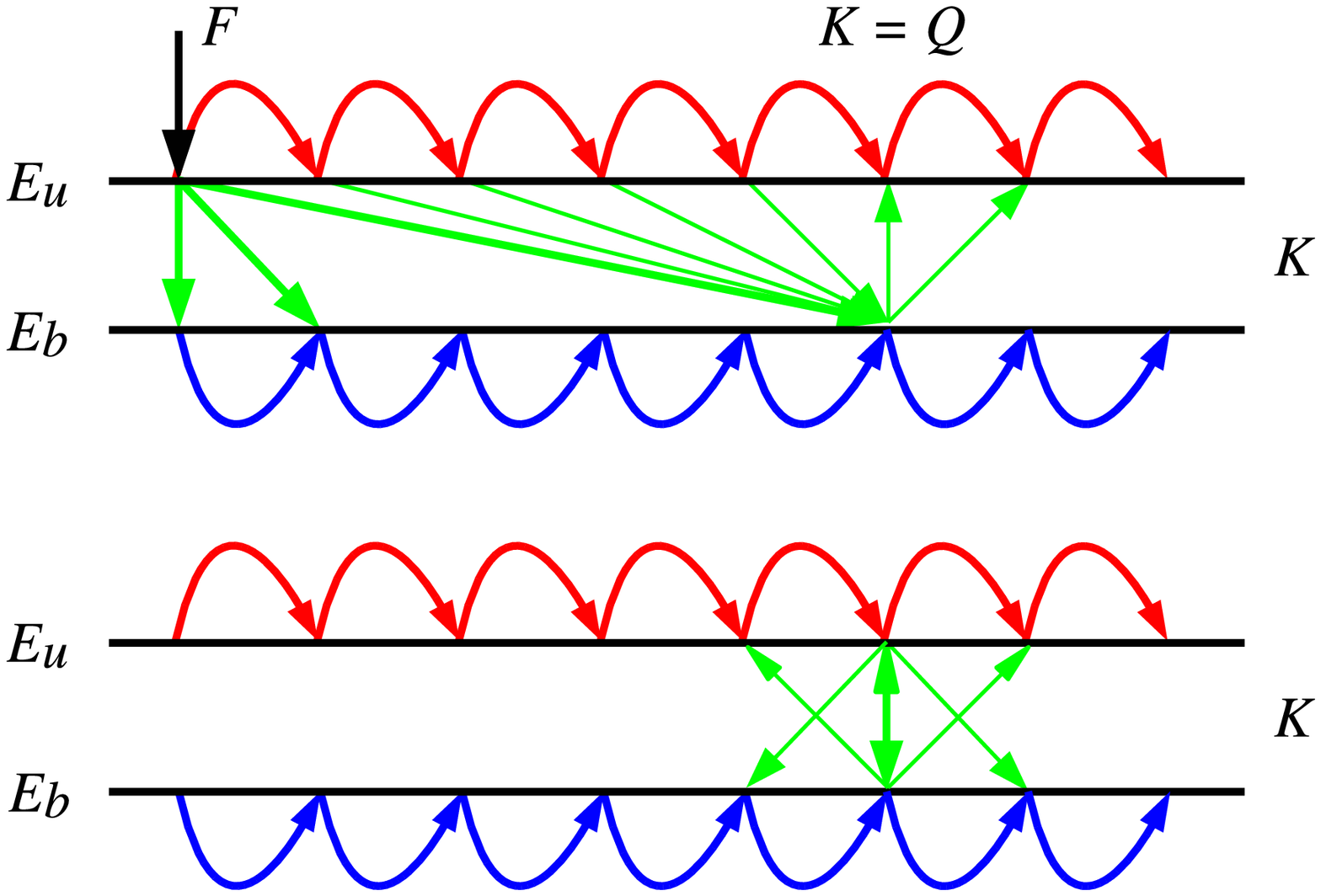}}
\caption{Sketch of the several shell-to-shell energy transfers identified 
in simulations of isotropic and homogeneous MHD turbulence. The $T_{uu}$ 
transfers are shown in red, $T_{bb}$ transfers are shown in blue, and 
$T_{ub}$ and $T_{bu}$ in green. The thickness of the arrows roughly 
indicates the strength of the transfers. Above: mechanically forced 
simulations. At the shell $K$, the magnetic field receives energy from 
the velocity field at all larger scales, and gives energy to the 
velocity field at slightly smaller scales. Below: Freely decaying 
turbulence. The $T_{ub}$ and $T_{bu}$ transfers only interchange energy 
between similar scales. In both cases, the $T_{uu}$ and $T_{bb}$ 
transfers are local and give the largest contribution to the flux.}
\label{fig6}
\end{figure}

The above considerations led several authors to consider whether 
some of the usual assumptions in hydrodynamic turbulence hold in the 
MHD case. From the shell-to-shell transfer, the scenario pictured in 
Fig. \ref{fig6} seems to arise for the energy: 
interactions between the same fields are 
mostly local, and interactions between the velocity and the magnetic field 
can have different degrees of non-locality depending on whether the 
turbulence is forced or freely decaying, depending on how the velocity 
and the magnetic fields are maintained against dissipation in the forced 
case, and depending on the presence of an external magnetic field. It is 
unclear for the moment whether the varying degree of non-locality with the 
configuration will converge to a universal solution for very large Reynolds 
numbers.

Theoretical arguments considering interactions in MHD turbulence also 
obtained conflicting results. Using the EDQNM closure, \citet{Pouquet76} 
reported non-local interactions which were associated to Alfv\'en waves. 
In \citet{Verma03,Verma04} and \citet{Verma05}, field-theoretic calculations 
were used to compute the shell-to-shell transfers and it was 
concluded that they were 
local except for the transfer between the velocity and the magnetic field, 
which was found to be somewhat non-local. The helicity transfer was also 
found to be non-local. Recently, \citet{Aluie10} gave strict bounds for 
fluxes in MHD turbulence under the assumptions that both the velocity and 
the magnetic energy follow power laws in the inertial range between 
$k^{-1}$ and $k^{-3}$. The velocity-to-velocity and 
magnetic-to-magnetic fluxes were found to be local in the limit of infinite 
Reynolds number, and the fluxes coupling velocity and magnetic fields were 
found to be local although counterexamples to their proof as the ones 
mentioned in Sect. 2 were acknowledged. However, these results shed 
light into why some simulations were found to be more local than others, 
as mechanisms as the small-scale dynamo can be expected to be less 
relevant in freely decaying turbulence in approximate equipartition 
between the two fields.

At presently attainable spatial resolutions, 
other indications of possible non-universal behavior has been reported 
in numerical simulations. In \citet{Dmitruk03}, simulations of 
forced reduced magnetohydrodynamics (RMHD) where presented where the 
energy spectrum changed 
its power law depending on the timescale of the external forcing. 
Spectra compatible with Kolmogorov, Iroshnikov-Kraichnan, weak turbulence 
theory, or even steeper laws were observed. 
The RMHD equations correspond to an 
approximation of the MHD equations when a strong external magnetic field 
is imposed. Similar results were reported by \citet{Mason08}, who considered 
forced MHD with an imposed magnetic field. Other numerical simulations 
of forced MHD turbulence (see e.g., 
\citealt{Muller03,Haugen03,Muller05,Beresnyak09}) 
also reported conflicting results. In freely decaying isotropic turbulence, 
some simulations were observed to develop Iroshnikov-Kraichnan scaling while 
others Kolmogorov-like scaling \citep{Muller05,Mininni07b,Mininni09}. 
Recently, large resolution simulations 
of freely decaying MHD flows showed that 
depending on the amplitude of the 
dynamically consistent large-scale magnetic field, different 
power laws can be realized \citep{Lee09}. Finally, recent studies of 
spectral laws in solar wind data \citep{Podesta07} indicate that many of 
these power laws can also be identified in space plasmas.

Although the main aim of this review is to consider studies of scale 
interactions in MHD, in this context it is worth mentioning some of the 
existing phenomenological theories for MHD turbulence. While Iroshnikov 
and Kraichnan considered small scale fluctuations as isotropic, it is clear 
now that MHD turbulence does not recover isotropy at small scales 
\citep{Shebalin83,Goldreich95,Milano01} and may become even more anisotropic 
as the scales are decreased. To take this into account, a different MHD 
spectrum has been advocated in \citet{Goldreich95}, whereby the anisotropy 
of the flow induces a Kolmogorov-like spectrum in the perpendicular direction. 
A balance between linear and non-linear timescales (the Alfv\'en and turnover 
times) is assumed which leads to a ``critical balance'' of the form 
$k_\parallel B_0 \sim k_\perp b_l$. Another anisotropic model based on 
dynamic alignment of the velocity and magnetic fields \citep{Boldyrev06} 
gives IK-like scaling in the perpendicular direction. In this case, the 
angle between the two fields decreases (and therefore the fields become 
more aligned) with the scale as $\sim l^{1/4}$. Consideration of this 
alignment in the Politano-Pouquet relations leads to the aforementioned 
scaling for the energy spectrum. Early extensions to flows with 
sizable cross-correlations can be found in \citet{Grappin83,Galtier00}. 
Other models have considered transitions from Kolmogorov 
to Iroshnikov-Kraichnan scaling by taking 
different combinations of the non-linear and Alfv\'en timescales 
\citep{Matthaeus89}, or consider non-locality \citep{Alexakis07c}.

Therefore, while the assumptions of locality and of isotropization of the 
small scales common in hydrodynamic turbulence allow for a simpler 
phenomenological treatment of MHD, the development of local anisotropies, 
the variety of time scales in the problem (see \citealt{Zhou04} for a 
review), and the different simulations showing scaling consistent with 
different phenomenological theories, led some authors to discuss some 
of these assumptions. In \citet{Schekochihin08}, non-locality, anisotropy, 
and non-universality were considered as defining properties of MHD 
turbulence. The authors argued that the small-scale dynamo, a fundamental 
process in MHD turbulence, shows clear signatures of non-locality 
\citep{Schekochihin02,Schekochihin02b,Haugen03,Haugen04,Schekochihin04,Mininni05b}. 
They also argued that anisotropy is intrinsic to MHD, and that 
non-universality manifests itself just from the needed distinction 
between MHD turbulence in the presence and in the absence of a strong mean 
field. Similar concerns about universal behavior in MHD were discussed in 
\citet{Lee09} for the case of freely decaying turbulence. \citet{Beresnyak09} 
consider the lack of a ``bottleneck'' in MHD (an accumulation of 
energy at the beginning of the viscous range observed in hydrodynamic 
turbulence) as evidence of non-locality (see also \citealt{Pietarila09}). 
In some sense, 
some of these discussions can be tracked back to early considerations of 
freely decaying MHD turbulence and the processes of selective decay 
\citep{Matthaeus80,Ting86,Kinney95,Mininni05e} and dynamic alignment 
\citep{Grappin83,Pouquet86,Ghosh88,Mininni05a}. MHD, having three ideal 
invariants, is known to decay for very long times into different attractors 
depending on the initial ratio of these invariants \citep{Ting86,Stribling91}. 
Although these solutions involve final stages of the decay, recent numerical 
simulations showed that these relaxed states can be realized locally in the 
flow in very short times scales 
\citep{Mason06,Matthaeus08,Servidio08,Perez09}, 
giving rise to different regimes.

\section{Concluding remarks}

Since the success of the phenomenological theory of Kolmogorov in 
hydrodynamic turbulence, several attempts have been made to apply similar 
considerations to tackle MHD turbulence. The presence of waves and 
of several time scales, and of several ideal invariants limited these 
approaches giving rise to many possible models. Solar wind observations 
and numerical simulations later showed that assumptions like isotropy 
of the small scales, or equipartition between the fields, may not hold 
in the MHD case. More recently, the increase in computing power allowed 
for some exploration of the parameter space giving rise to conflicting 
results for scaling laws in the energy spectrum.

The recent introduction of shell-to-shell transfers allowed for detailed 
studies of scale interactions in MHD turbulence and opened the door to 
discuss another hypothesis: that of locality of interactions between 
scales. The results, at intermediate spatial resolutions and Reynolds 
numbers, show different degrees of non-locality depending on the 
configuration studied: forced or freely decaying turbulence, in presence 
or in absence of an external magnetic field, etc. Non-local transfers, 
when observed, involve the coupling between the velocity and the magnetic 
field, or the transfer of magnetic helicity. In the former case, the 
non-local transfers were not larger than $10-20\%$ of 
the total, although they played fundamental roles, e.g., sustaining 
the magnetic field by dynamo action against Ohmic dissipation.

In spite of some conflicting results in the simulations and theory, 
there is growing consensus that MHD turbulence is less local than 
hydrodynamic turbulence. To what extent it is a matter of debate. 
It is unclear for the moment whether these effects will go away for 
larger Reynolds numbers, or if they stay how much impact they will have 
in the flow dynamics, and under what conditions. However, the different 
degrees of non-locality observed at present resolutions, and the 
existence of non-local processes in MHD (as, e.g., the small scale dynamo), 
call for a discussion about the validity of the hypothesis of locality 
of interactions, and of whether there is only one kind of MHD turbulence 
or many. This raises the question of what is the definition of MHD 
turbulence in phenomenological or theoretical approaches. If only 
configurations as the ones in the solar wind (with an imposed magnetic 
field) are to be considered, then a universal scaling (or several 
classes of universality) may be possibly identified. However, if 
processes like the small-scale dynamo, the large-scale dynamo, and 
inverse cascades are to be considered as manifestations of MHD 
turbulence, non-local interactions and non-universal behavior may 
persist even for very large Reynolds numbers. In this context, many 
of the works reviewed here may have to be revisited in the following 
years, as experiments and increased computing power will allow us to 
explore new regions of the parameter space of MHD turbulence.

\section{Summary points}

\begin{enumerate}
\item  Simply applying properties of hydrodynamic turbulence to the 
MHD case may not be possible. In particular, assumptions of scale 
locality of MHD turbulence must be tested in experiments and simulations.

\item Shell-to-shell transfer functions allow for detailed studies of 
coupling between fields and scales in numerical simulations. The 
shell-to-shell transfers can also be associated with physical processes 
such as as Alfv\'en wave interactions, Joule damping, and dynamo action.

\item The degree of non-locality observed at the presently attainable 
spatial resolutions depends on the configuration.

\item Mechanically forced turbulence shows local transfer of magnetic 
and kinetic energy, but the coupling between the velocity and magnetic 
field that sustains the latter against Ohmic dissipation is non-local.

\item In freely decaying MHD turbulence, non-local effects seem to be 
negligible.

\item Studies of the energy transfer in the presence of an imposed 
magnetic field show most of the transfer takes place in the direction 
perpendicular to the external field, with strong non-local interactions 
with modes with $k_\parallel=0$.

\item The transfer of energy for the Els\"asser variables is more local 
than the transfer in terms of the velocity and magnetic fields.

\item The shell-to-shell transfer of magnetic helicity is more complex, 
with superimposed direct and inverse transfers. The inverse transfer 
has a local component, and a non-local one that moves energy from the 
forced scale directly to the largest scales in the system.

\end{enumerate}

\section{Acknowledgments}

The National Center for Atmospheric Research is sponsored by the National 
Science Foundation. The author acknowledges fruitful discussions with 
Alexandros Alexakis, Daniele Carati, Gregory Eyink, Annick Pouquet, and 
Alexander Schekochihin, and figures gently provided by A. Alexakis 
and D. Carati. The author also expresses his gratitude to A. Pouquet 
for her careful reading of the manuscript. PDM is a member of the Carrera 
del Investigador Cient\'{\i}fico of CONICET, and acknowledges support from 
grants UBACYT X468/08 and PICT 2007-02211.

\bibliography{ms}

\end{document}